\documentclass[aps,pre,twocolumn,superscriptaddress]{revtex4-1}
\usepackage{graphicx}
\usepackage{amsmath,amsthm,amssymb,physics,color,enumerate,tabularx,makecell,mathrsfs,bbm,stackrel,mathtools,multirow,soul}
\begin{document}

\title{Role of topology in determining the precision of a finite thermometer}
\author{Alessandro Candeloro}
\email{alessandro.candeloro@unimi.it} 
\affiliation{Quantum Technology Lab, Dipartimento di 
Fisica {\em Aldo Pontremoli}, Universit\`{a} degli Studi di Milano, I-20133 Milano, Italy}
\author{Luca Razzoli}
\email{luca.razzoli@unimore.it}
\affiliation{Dipartimento di Scienze Fisiche, Informatiche e 
Matematiche, Universit\`{a} di Modena e Reggio Emilia, I-41125 Modena, 
Italy}
\author{Paolo Bordone}
\email{paolo.bordone@unimore.it}
\affiliation{Dipartimento di Scienze Fisiche, Informatiche e 
Matematiche, Universit\`{a} di Modena e Reggio Emilia, I-41125 Modena, 
Italy}
\affiliation{Centro S3, CNR-Istituto di Nanoscienze, I-41125 Modena, Italy}
\author{Matteo G. A. Paris}
\email{matteo.paris@fisica.unimi.it}
\affiliation{Quantum Technology Lab, Dipartimento di 
Fisica {\em Aldo Pontremoli}, Universit\`{a} degli Studi di Milano, I-20133 Milano, Italy}
\affiliation{INFN, Sezione di Milano, I-20133 Milano, Italy}

\begin{abstract}
Temperature fluctuations of a finite system follows the Landau bound $\delta T^2 = T^2/C(T)$ where $C(T)$ is the heat capacity of the system. In turn, the same bound sets a limit to the precision of temperature estimation when the system itself is used as a thermometer. In this paper, we employ graph theory and the concept of Fisher information to assess the role of topology on the thermometric performance of a given system. We find that low connectivity is a resource to build precise thermometers working at low temperatures, whereas highly connected systems are suitable for higher temperatures. Upon modeling the thermometer as a set of vertices for the quantum walk of an excitation, we compare the precision achievable by position measurement to the optimal one, which itself corresponds to energy measurement.
\end{abstract}
\date{\today}
\maketitle

\section{Introduction}
\label{sec:intro}
Thermometry is based on the zero-th law of thermodynamics. A probing object (the thermometer) is put in contact with the system under investigation and when they achieve thermal equilibrium the temperature of both is determined by performing a measurement on the thermometer. Good thermometers are those with a heat capacity much smaller than the object under study, such that the thermal equilibrium is reached at a temperature very close to the original temperature of the object. This straightforward line of reasoning leads to consider {\em small} thermometers, possibly subject to the laws of quantum mechanics \cite{revmes}. Additionally, since the heat capacity itself depends on temperature, one is led to investigate whether the heat capacity of a thermometer may be tailored for a specific range of temperatures \cite{cth}.
\par
The topic has become of interest in the last two decades, due to the development of controlled quantum systems at the classical-quantum boundary \cite{par89,jar97,muk03,sei05,cuc92,ple98,Nie02,jac12,
gha14,ali14,pla14,bra15,bin15,bru15,bor15,esp15,ols15,pek15}, which makes it 
relevant to have a precise
determination of temperature for quantum systems
\cite{nmr0,nmr1,nmr7,nmr8,nmr9,nmr10,nmr11}, and to 
understand the ultimate bounds to precision in the estimation of 
temperature 
\cite{bru06,sta10,bru11,bru12,mar13,hig13,cor14,meh15,jev15,jar15,adp15,adp17}.
At the same time, precise manipulation of quantum systems makes it possible to design and realize quantum thermometers, i.e., thermometers where temperature is precisely estimated looking at tiny changes in genuine quantum features such as entanglement or coherence \cite{sho19,sho20,mit20,pot20}.
\par
As a matter of fact, temperature is not an observable in a strict sense, i.e., it is not possible to build a self-adjoint operator corresponding to temperature. Besides, temperature represents a macroscopic manifestation of random energy exchanges between particles and, as such, does 
fluctuate for a system at thermal equilibrium. In fact, this has made the concept of 
temperature fluctuations controversial
\cite{Man6x,mcf73,Kit73,Kit88,Man89,Pro93,Chu92,Bol10,uff99}.  In order to retain
the operational definition of temperature, one should conclude that although 
temperature itself does not fluctuate, any temperature 
estimate is going to fluctuate, since it is based on the measurement of one or more 
proper observables of the systems, e.g., energy or population.
\par
In this framework, upon considering temperature as a function 
of the exact and fluctuating values of the other state parameters, 
Landau and Lifshitz derived a relation for the temperature fluctuations 
of a finite system\cite{LanLi,Phi84}.  This is
given by $ \delta T^2 = T^2/C$ where $C=C(T)$ is the (temperature-dependent) 
heat capacity of the system and appears as a fundamental bound to the precision
of any temperature estimation. The same problem may be addressed by leveraging 
tools from quantum parameter estimation and the Landau bound may be shown to be 
equivalent to the so-called Cram\'er-Rao bound to precision, built by evaluating 
the quantum Fisher information (QFI) of equilibrium states \cite{paris2015achieving}.
In turn, the link between the QFI and the heat capacity have been established in different frameworks, such as in quantum phase transition and in systems with vanishing gap \cite{liu2019quantum,zanardi2007bures,paris2015achieving,zanardi2007information,Potts2019fundamentallimits}.
\par
In this paper, we exploit the above connection to address the role of topology in 
determining the precision of a finite thermometer. In particular, upon modeling 
a finite thermometer as a set of connected subunits, we employ graph theory, together with 
QFI, to assess the role of topology on the thermometric 
performance of the system. We confirm that measuring the energy of the system is the best 
way to estimate temperature, and also find that systems with 
low connectivity are suitable to build precise thermometers working at 
low temperatures, whereas highly connected systems are suitable for higher temperatures. 
We also compare the optimal precision with that achievable by measuring the 
position of thermal excitations. Our results indicate that quantum probes are especially 
useful at low temperatures and that systems with low connectivity provide more precise 
thermometers. At high temperatures, precision degrades as $O(T^{4})$ with highly
connected systems providing at least a better proportionality constant. Reference models are physical systems in which the connectivity plays a relevant role, e.g. quantum dots arranged in lattices \cite{baimuratov2013quantum} and qubits in quantum annealers \cite{lechner2015quantum,nigg2017robust}. Evidences suggest that a system of qubits in D-Wave quantum annealers quickly thermalizes with the cold environment \cite{buffoni2020thermodynamics} and that a pause mid-way through the annealing process  increases the probability of successfully finding the ground state of the problem Hamiltonian, and this has been related to the thermalization of the system \cite{marshall2019power}.
\par
The paper is structured as follows. In Sec. \ref{sec:equilibriumThermo} we briefly review the tools of quantum estimation theory, focusing on the ultimate performance of equilibrium states in estimating the temperature of an external environment. According to these results, in Sec. \ref{sec:networkThermo} we consider the equilibrium states of the Laplacian matrix of simple graphs, addressing the efficiency of our probes in both the high- and low- temperature regimes. The Laplacian matrix is indeed the Hamiltonian of a quantum walker moving on discrete positions. In Sec. \ref{sec:results} we derive analytical and numerical results for some remarkable simple graphs and two-dimensional lattices. In Sec. \ref{sec:conclusion} we summarize
and discuss our results and findings. Then, in the appendixes we offer analytical proofs and details of the results presented.

\section{Equilibrium  Thermometry}
\label{sec:equilibriumThermo}
\subsection{Estimation Theory}
Given an experimental set of outcomes of size $M$ $\{\vec{x}\}\in\mathcal{M}^{\oplus M}$ which depends on some parameter $\lambda$, we can infer the value of the parameter through an estimator function $\hat{\lambda}(\vec{x})$. The variance $\textup{Var}(\hat{\lambda})$ is the usual figure of merit that quantifies the precision of an estimator: the lower the variance, the closer the outcomes are spread around the expected value of the estimator. According to the probability distribution $p(x_m\vert \lambda)$ of the outcomes (throughout the section we will consider a discrete set of outcomes $\mathcal{M}$ with cardinality $N_{\mathcal{M}}$), the variance of any unbiased estimators of the parameter $\hat{\lambda}(\vec{x})$ can be lower bounded as 
\begin{equation}
\label{eq:crb}
    \textup{Var}(\hat{\lambda}) \geq \frac{1}{M \mathcal{F}_c(\lambda)}\,,
\end{equation}
where 
\begin{equation}
\label{eq:fi}
    \mathcal{F}_c(\lambda) = \sum_{m=1}^{N_{\mathcal{M}}} \frac{(\partial_\lambda p(x_m\vert \lambda))^2}{p\textbf{}(x_m\vert \lambda)}\,.
\end{equation}
In the literature, this result is known as the Cram\'{e}r-Rao bound (CRB) \cite{lehmann2006theory,van2004detection}, and $\mathcal{F}_c(\lambda)$ is the Fisher information (FI) for the statistical model $p(x_m\vert\lambda)$. The latter quantifies how much information on $\lambda$ is encoded in the probability distribution: a large FI means that the outcomes carry significant information on the parameter, which is reflected by the possibility of having more precise estimators, see Eq. \eqref{eq:crb}. The attainability of the CRB is the fundamental problem of classical estimation theory. Indeed, it is known that the lower bound can be saturated by the maximum likelihood estimator in the limit of infinite set of measurements $M\to+\infty$ \cite{newey1994large}.
\par
If we move to the quantum realm, observables are described by self-adjoint operators. However, if the quantity of interest is not an observable {(such as the temperature)}, then we can not directly measure it. For this reason, one needs the tools provided by quantum estimation theory to find the best optimal probing strategy. As known, in quantum mechanics probability distributions are naturally described by the Born rule $p(x_m\vert \lambda) = \Tr[\rho_\lambda \Pi_{m}]$, in which we have assumed that the information of the parameter is encoded in the density matrix, while the measurement is $\lambda$-independent and it is identified by the set of positive operator-valued measures $\{\Pi_m\}_m$. From this perspective, we see that there is arbitrariness in the choice of the positive operator-valued measure (POVM). Thus, once the state $\rho_\lambda$ is fixed, we have a family of possible probability distribution depending on this choice. Among the FI arising from all the possible POVMs, we can show that \cite{paris2009quantum} there is an optimal POVM that maximizes the FI, which is given by the set of projectors $\{\vert L_j\rangle\langle L_j\vert\}_j$ of the symmetric logarithmic derivative $\Lambda_\lambda$, implicitly defined as
\begin{equation}
    2 \partial_\lambda \rho_\lambda = \hat{\Lambda}_\lambda\rho_\lambda + \rho_\lambda \hat{\Lambda}_\lambda.
\end{equation}
The maximum of the FI among all the possible POVMs is known as the quantum Fisher information (QFI) \cite{amari2007methods}, which can be obtained as
\begin{equation}
    \mathcal{F}_c(\lambda) \leq \mathcal{F}_q(\lambda) = \Tr[\rho_\lambda \hat{\Lambda}_\lambda^2].
\end{equation}
From that, we have a corresponding quantum inequality for the variance of any estimator  which is known as the quantum Cram\'{e}r-Rao bound (CRB) \cite{PhysRevLett.72.3439}
\begin{equation}
\textup{Var}({\hat{\lambda}}) \geq \frac{1}{M\mathcal{F}_c(\lambda)} \geq \frac{1}{M\mathcal{F}_q(\lambda)}.
\end{equation}
Thus, the QFI sets the minimum attainable error among the sets of all probing schemes in the estimation problem of $\lambda$. Notice that all these considerations hold as long as the state $\rho_\lambda$ is fixed. 

\subsection{Quantum Fisher Information}
In this paper we focus on a finite-size quantum system living in a $N$-dimensional Hilbert space and described by a Hamiltonian operator  $\hat{H} = \sum_k E_k \vert e_k\rangle\langle e_k\vert$, with $k=0,...,N-1$. The idea is to use a finite system as a probe to estimate the temperature $T$ of an external environment. We thus consider the customary thermodynamic situation occurring in thermalization processes, when a system is in contact with a thermal bath at temperature $T$ and, after some time, it eventually reaches an equilibrium state at the same temperature $T$ of the bath.  The final equilibrium state of the probing system is thus given by the Gibbs state
\begin{equation}
\rho_T = \frac{1}{Z} e^{-\hat{H}/T} = \sum_n\sum_{\alpha=1}^{g_n} \frac{e^{-E_n/T}}{Z} \vert e_{n,\alpha}\rangle\langle e_{n,\alpha} \vert\,.
\label{eq:equilstate}
\end{equation}
Throughout the paper we set the Boltzmann constant $k_B=1$. In the last equality we make explicit the possible degeneracy of the energy levels:
$n$ labels the distinct energy levels, and $g_n$ is the corresponding degeneracy. In terms of the latter, the partition function $Z$ can be written as
\begin{equation}
    Z = \sum_{k=0}^{N-1} e^{-E_k/T} = \sum_n g_n e^{-E_n/T}.
\end{equation}

Since the state \eqref{eq:equilstate} is diagonal in the energy eigenbasis, and since the latter does not depend on the parameter $T$, the statistical model reduces to a classical-like estimation problem, where the optimal POVM is realized exactly by $\{\vert e_{n,\alpha}\rangle \langle e_{n,\alpha}\vert\}$. Moreover, the QFI is easily obtained and turns out to be proportional to the variance of the Hamiltonian operator $\hat{H}$, i.e.
\begin{equation}
    \mathcal{F}_q(T) = \frac{1}{T^4}\left(\langle \hat{H}^2\rangle - \langle \hat{H}\rangle^2\right)\,,
    \label{eq:qfi_def}
\end{equation}
where the expectation values of $\hat{H}^p$ for a Gibbs state are given as
\begin{equation}
\langle \hat{H}^p \rangle = \sum_n g_n \frac{e^{-E_n/T}}{Z}E_n^p\,.
\end{equation}

\subsection{Fisher Information for a position measurement}
Our system lives in a $N$-dimensional space, and we assume the position space to be finite and discrete. When a system is confined to discrete positions, a position measurement is a suitable and standard measurement. In this section we study how informative the position measurement is for estimating the temperature.  The POVM is given by $\{\vert j \rangle \langle j \vert\}$, where $j=0,\ldots,N-1$ labels the discrete positions.
The probability of observing the system in the $j$th position given the temperature $T$ is 
\begin{equation}
    p(j\vert T ) = \Tr[\rho_T \vert j \rangle \langle j \vert] = \sum_{k=0}^{N-1} \frac{e^{-E_k/T}}{Z} \vert \langle j \vert e_k \rangle \vert^2. 
\end{equation}
Therefore, the FI \eqref{eq:fi} for the position measurement is
\begin{equation}
\mathcal{F}_c(T) = \sum_{j=0}^{N-1} \frac{(\partial_T p(j\vert T))^2}{p(j\vert T)}\,,
\label{eq:FI_pos_def}
\end{equation}
which can be rewritten (see Appendix \ref{app:FIposition}) in a form similar to that of the QFI,
\begin{equation}
\label{eq:fisherinfoposition}
\mathcal{F}_c(T) = \frac{1}{T^4} \left(\sum_{j=0}^{N-1}\frac{\langle \hat{H}\rho_T\rangle_j^2}{p(j\vert T)} - \langle \hat{H}\rangle^2\right),
\end{equation}
where 
\begin{equation}
\label{eq:EnergyWeighted}
\langle \hat{H}\rho_T \rangle_j = \sum_{k=0}^{N-1} \frac{e^{-E_k/T}E_k}{Z}  \vert \langle j \vert e_k \rangle \vert^2
\end{equation}
is the expectation value of $\hat{H}\rho_T$ on the position eigenstate $\vert j \rangle$.
\par

\section{Network Thermometry}
\label{sec:networkThermo}
We focus on the estimation of temperature using quantum probes which may be regarded as set of connected subunits, i.e., described by connected simple graphs (undirected and not multigraph). A graph is a pair $G=(V,E)$ where $V$ denotes the nonempty set of vertices and $E$ the set of undirected edges, which tell which vertices are connected. The set of vertices is the finite set of discrete positions the quantum system can take. The set of edges accounts for all and only the possible paths the system can follow to reach two given vertices. The number of vertices $\vert V \vert = N$ determines the order of the graph, and the number of the edges is $\vert E \vert = M$.  All this information determines the topology of the graph and is encoded in the Laplacian matrix $L = D - A$. In the position basis $\{\vert j \rangle \langle j \vert \}$, the degree matrix $D$ is diagonal with elements $D_{jj} = \textup{deg}(j)=:d_j$, the degree of vertex $j$, while the adjacency matrix $A$ has elements $A_{jk} = 1$ if the vertices $j$ and $k$ are connected by an edge or $A_{jk}=0$ otherwise. A graph $G$ is said to be $k$-regular if all its vertices have the same degree $k$. The Laplacian matrix for an undirected graph is positive semidefinite, and symmetric, and the smallest Laplacian eigenvalue is $0$, which, for connected graphs, has degeneracy $g_0 = 1$. Instead, the second-smallest Laplacian eigenvalue is also known as algebraic connectivity \cite{de2007old,alavi1991laplacian,wu2005algebraic}: smaller values represent less connected graphs.
\par
A continuous-time quantum walk (CTQW) is the motion of a quantum particle with kinetic energy when confined to discrete positions, e.g. the vertices of a graph. In a discrete-position space, $-\nabla^2$ of the kinetic energy is replaced by the Laplacian matrix $L$ \cite{wong2016laplacian}. Hence, a quantum walker has intrinsically the topology of the graph, and so it is a promising candidate to be the probe for estimating the temperature of an external environment with respect to the topology of the network. The CTQW Hamiltonian is $H = \gamma L$, where the parameter $\gamma > 0$ is the hopping amplitude of the walk and accounts for the energy scale of the system. We have already set $k_B = 1$, and in the following we also set $\gamma = 1$. Therefore, energy and temperature are hereafter dimensionless. Notice that, in this way, the energy eigenvalues $E_n$ are the Laplacian eiegenvalues and the scale of temperature should be intended as referred to the energy scale specific of the system considered, i.e., to $\gamma$.

\begin{figure}[!htb]
    \centering
    \includegraphics[width=0.48\textwidth]{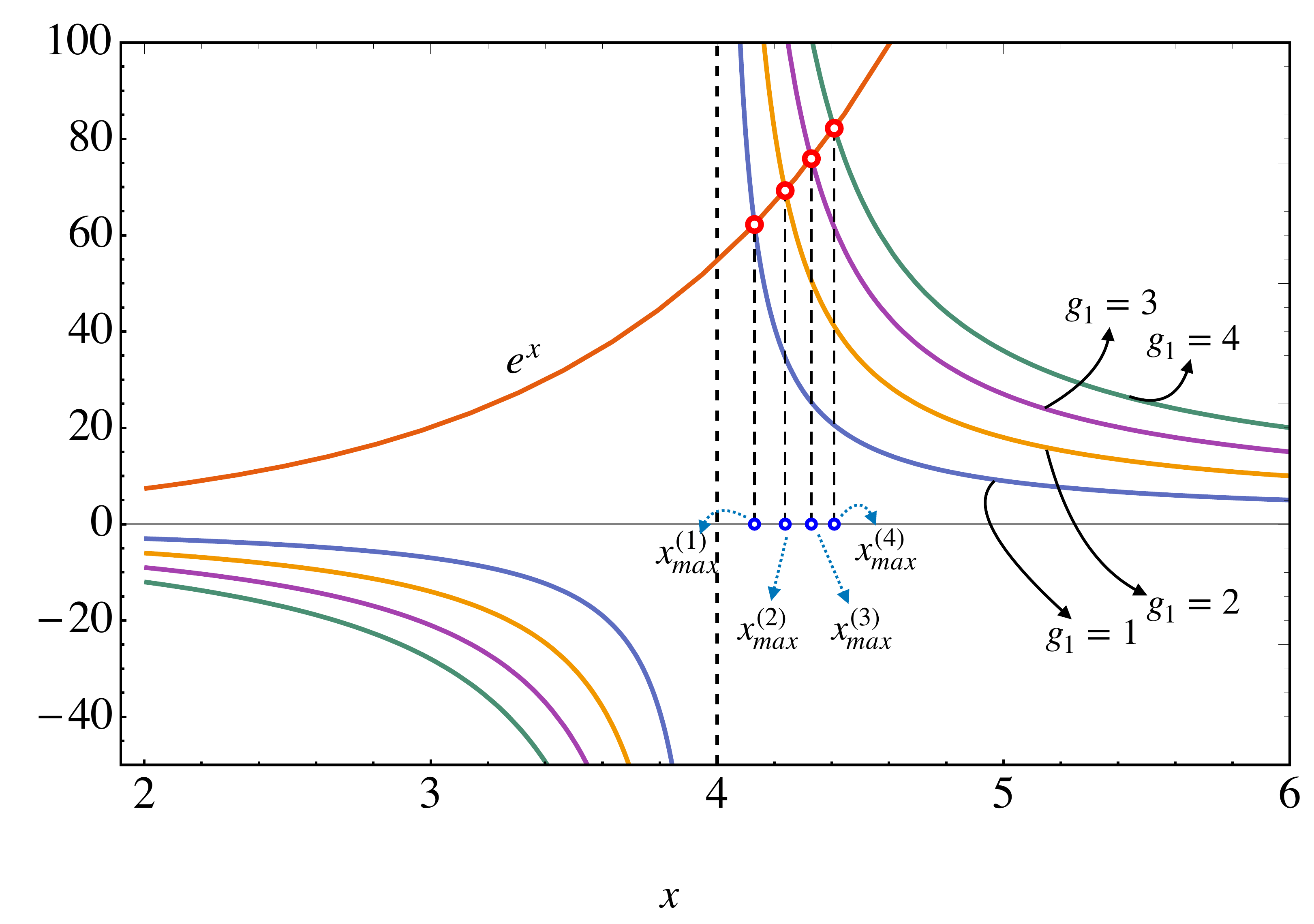}
    \caption{Graphical solution $x_{max}^{(g_1)}$ of the transcendental equation \eqref{eq:transcendental} for different values of $g_1$.}
    \label{fig:graph_sol_maxQFI}
\end{figure}

\subsection{Low-temperature regime}
\label{subsec:lowTreg}
First, we analyze the regime of low temperatures $T$. Therefore, we assume that the system is mostly in the ground state and can access only the first excitation energy $E_1$, i.e. $E_n\gg T$ for $n>1$. For this reason, the partition function is
\begin{equation}
    Z = 1 + g_1 e^{-E_1/T}\,.
\end{equation}
Since the ground state energy is null, the mean value of the energy is $\langle \hat{H}\rangle =  g_1 E_1 e^{-E_1/T}Z^{-1}$ and it follows that the QFI in the low temperature regime can be approximated as
\par
\begin{equation}
    \mathcal{F}^{low}_q(x) \simeq \frac{f_{g_1}\left(x\right)}{E_1^2}\,,
\label{eq:lowTqfi_approx}
\end{equation}
where $x=E_1/T$ and the function $f_{g_1}(x)$ is defined as
\begin{equation}
f_{g_1}(x) = \frac{g_1 x^4 e^{-x}}{(1+g_1 e^{-x})^2}\,.
\end{equation}
The value $x_{max}>0$ at which the latter exhibits the maximum is the solution of the following transcendental  equation
\begin{equation}
    e^{x_{max}} = g_1 \frac{x_{max}+4}{x_{max}-4}\,,
    \label{eq:transcendental}
\end{equation}
obtained from $d f_{g_1}(x)/dx =0$. Solutions of this equation can be obtained graphically, as shown in Fig. \ref{fig:graph_sol_maxQFI}. The $x_{max}$ depends only on the degeneracy $g_1$ and numerical results show that $x_{max}(g_1)$ is a sublinear function, i.e. it increases less than linearly with $g_1$.

The FI in the low-temperature regime can be approximated as
 \begin{equation}
     \mathcal{F}^{low}_c(T) \simeq \frac{E_1^2 e^{-2E_1/T}}{Z T^4}
     \left( \sum_{j=0}^{N-1} \frac{\eta_j^2}{\frac{1}{N}+e^{-E_1/T}\eta_j}-\frac{g_1^2}{Z}\right)\,,
     \label{eq:lowTfi}
 \end{equation}
 since $p(j\vert T) \simeq \frac{1}{Z}\left(\frac{1}{N} + \exp{-E_1/T} \eta_j\right)$ and $\langle \hat{H}\rho_T \rangle_j \simeq \frac{1}{Z} \exp{-E_1/T}E_1 \eta_j$, where $\eta_j = \sum_{\alpha = 1}^{g_1} \vert \langle j \vert e_{1,\alpha} \rangle \vert ^2$.

\subsection{High-temperature regime}
\label{subsec:highTreg}
We move now to the opposite regime, high temperature, in which we assume that $T \gg E_k$ for all $k$.
The single-walker probe is no longer valid in the high-temperature regime, where many excitations, not only one, come into play. Yet, it can be used for small thermometers with bounded spectrum and large energy gap $E_1-E_0$, so that we may expect few excitations, and the single-walker model can still approximate the real system.
In this regime, the density matrix, in the energy eigenbasis, can be approximated by the maximally mixed state $\rho_T \simeq I_N/N$, where $I_N$ is the $N\times N$ identity matrix.
Accordingly, the QFI becomes
\begin{align}
	\mathcal{F}^{high}_q(T) & \simeq \frac{1}{T^4}\left[\frac{1}{N} \sum_{k=0}^{N-1} E_k^2 - \frac{1}{N^2}\left(\sum_{k=0}^{N-1} E_k\right)^2\right] \nonumber \\
    & = \frac{1}{N T^4} \left[ \sum_{k=0}^{N-1} d_k^2 + 2M\left(1-\frac{2M}{N}\right)\right]\,.
    \label{eq:QFI_highT}
\end{align}
Refer to Appendix \ref{app:qfihighT} for details on the sum of the energy eigenvalues and that of their square.
Thus, in the limit of high temperatures, the QFI tends to zero as $O(T^{-4})$ and proportionally to a topology-dependent factor.
\par
The sum of the squared degree can be bounded as
\begin{equation}
\label{eq:degreesquared}
\frac{4M^2}{N} \leq \sum_{k=0}^{N-1} d_k^2 \leq M \left(\frac{2M}{N-1} + N-2\right)\,,
\end{equation}
where the upper bound is proved in \cite{de1998upper} and the lower bound follows from the Cauchy-Schwartz inequality for the inner product of two $N$-dimensional vectors, $(1,\ldots,1)$ and $(d_0,\ldots,d_{N-1})$, using $\sum_{k=0}^{N-1} d_k=2M$. Hence, we can bound  $\mathcal{F}^{high}_q(T)$ as
\begin{equation}
\frac{2M}{NT^4} \leq \mathcal{F}_q^{high}(T) \leq \frac{M}{T^4}\ \left[1-\frac{2M(N-2)}{N^2(N-1)}\right]\,.
\label{eq:bounds_QFI}
\end{equation}
The upper bound in \eqref{eq:degreesquared} is saturated by the complete graph, while the lower bound is saturated, e.g, by the cycle graph and the complete bipartite graph whose partite sets have both cardinality $N/2$: hence, these bounds are actually achievable, and, accordingly, the bounds \eqref{eq:bounds_QFI} on the QFI are saturated by the above mentioned graphs (see Sec. \ref{sec:results} for details).
For high temperatures the optimal thermometer is the complete graph, which, among the simple graphs, has the maximum number of edges $M$. Notice also that the complete graph has the maximum energy gap, since $E_1-E_0=N$. Thus, unlike the low-temperature regime, in the high-temperature regime the graphs which perform better are those with high connectivity, in the sense of those with a high number of edges $M$.
\par
Recalling that in the high-temperature regime $\rho_T \simeq I_N/N$, we can approximate the FI as
\begin{align}
    \mathcal{F}^{high}_c(T) &\simeq \frac{1}{N^2 T^4} \left[N \sum_{j=0}^{N-1}\left( \sum_{k=0}^{N-1} E_k \vert \langle j \vert e_k \rangle \vert^2 \right)^2 -4M^2\right]\nonumber\\
    &=\frac{1}{N^2 T^4} \left(N \sum_{j=0}^{N-1}d_j^2 -4M^2\right)\,,
    \label{eq:FI_highT}
\end{align}
where the second equality follows from
\begin{align}
    \sum_{k=0}^{N-1} E_k \vert \langle j \vert e_k \rangle \vert^2 &= \sum_{k=0}^{N-1} \langle j \vert L \vert e_k \rangle \langle e_k \vert j \rangle\nonumber\\
    &= \langle j \vert L \vert j \rangle = d_j\,.
\end{align}
Therefore, the asymptotic value of the ratio $\mathcal{F}_c (T)/\mathcal{F}_q(T)$ is
\begin{align}
\lim_{T \to +\infty} \frac{\mathcal{F}_c (T)}{\mathcal{F}_q(T)}
& = \frac{N \sum_{k=0}^{N-1}d_k^2 -4M^2}{N\left[ \sum_{k=0}^{N-1} d_k^2 + 2M\left(1-\frac{2M}{N}\right)\right]} \nonumber
\\
& = \frac{1}{1+\lambda_{N,M}}\,,
\label{eq:limit_ratio_FIQFI}
\end{align}
where we have introduced the quantity
\begin{equation}
    \lambda_{N,M} = \frac{2M}{\sum_{k=0}^{N-1}d_k^2-\frac{4M^2}{N}}
\end{equation}
to capture the (asymptotic) discrepancy between the FI and the QFI in terms of the topology features of the graphs: small $\lambda_{N,M}$ means a ratio close to 1, $\mathcal{F}_c (T)\simeq \mathcal{F}_q(T)$; large $\lambda_{N,M}$ means a ratio close to 0, $\mathcal{F}_c (T) \ll \mathcal{F}_q(T)$.

\subsection{Fisher Information for circulant graphs}
In this section we prove that the FI for position measurement is identically null in the case of circulant graphs, e.g., the complete graph and the cycle graph. A circulant graph is defined as the regular graph whose adjacency matrix is circulant, and accordingly so is the Laplacian matrix \cite{elspas1970graphs,golin2004unhooking,WolframCirculantGraph}.
A circulant matrix is a special Toeplitz matrix where every row of the matrix is a right cyclic shift of the row above it. The eigenproblem for circulant matrices is solved \cite{gray2006toeplitz}, and the Laplacian eigenstates of circulant graphs are
\begin{equation}
\label{eq:eigenvector_circulant}
    \vert e_k \rangle = \frac{1}{\sqrt{N}} \sum_{j=0}^{N-1} \omega^{kj} \vert j \rangle\,,
\end{equation}
with $\omega = \exp{2\pi i /N}$ and $0 \leq k \leq N-1$. This means that $\vert \langle j \vert e_k\rangle\vert^2 = 1/N\, \forall k$ and consequently
\begin{equation}
    p(j\vert T) =  \frac{1}{N}\,,
\end{equation}
while 
\begin{equation}
    \langle \hat{H}\rho_T\rangle_j = \frac{1}{N} \langle \hat{H}\rangle\,.
\end{equation}
From Eq. \eqref{eq:fisherinfoposition} we clearly see that $\mathcal{F}_c(T) =0$. We conclude that for circulant graphs the position measurement does not carry any information on the temperature $T$. 
\par
Actually, the result is more general: the FI for a position measurement is null not only for circulant graphs, but for all the graphs such that $\vert \langle j \vert e_k\rangle \vert^2 = t_j$ does not depend on $k$. Indeed, in this case we have $p(j\vert T) = t_j$ and $\langle \hat{H}\rho_T\rangle_j = t_j \langle \hat{H}\rangle$, from which we see that \eqref{eq:fisherinfoposition} is identically $0$, since $\sum_{j=0}^{N-1}t_j = 1$.

\begin{figure}[!htb]
\centering
\includegraphics[width =\columnwidth]{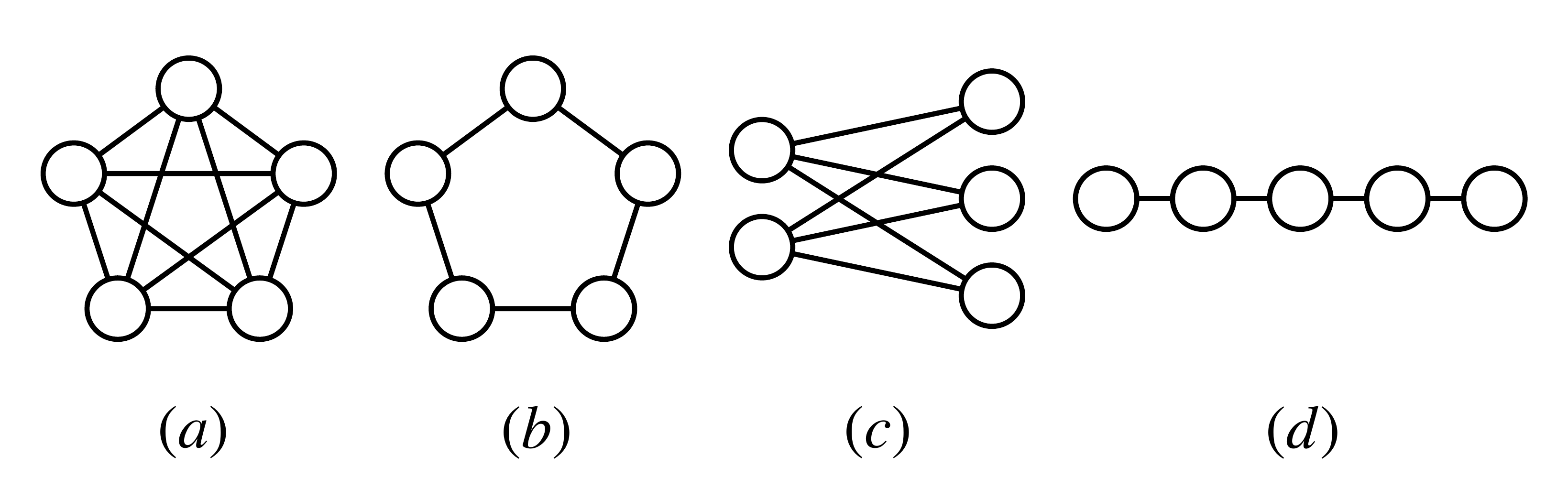}
\caption{Graphs considered in the present work (example for $N=5$ vertices): (a) Complete graph $K_5$, (b) cycle graph $C_5$, (c) complete bipartite graph $K_{2,3}$, and (d) path graph $P_5$.}
\label{fig:graphs}
\end{figure}

\section{Network Thermometry: results}
\label{sec:results}
\begin{figure*}
    \centering
    \includegraphics[width=\textwidth]{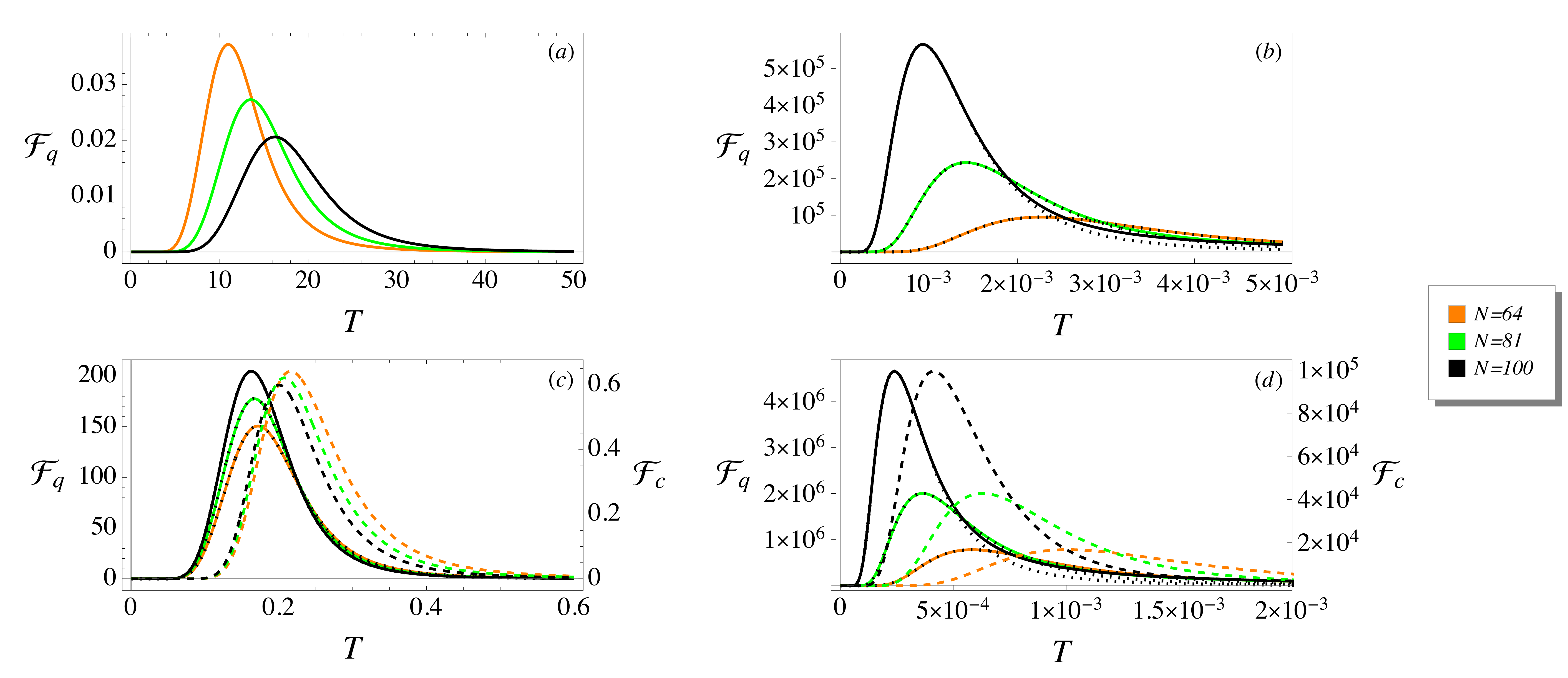}
    \caption{QFI and FI for position measurement for different graphs of order $N$: (a) complete graph, (b) cycle graph, (c) star graph, and (d) path graph. Solid colored line: QFI $\mathcal{F}_q$. Dotted black line: QFI at low temperature $\mathcal{F}_q^{low}$ \eqref{eq:lowTqfi_approx} (not reported for the complete graph since it coincides with Eq. \eqref{eq:QFI_complete}). Dashed colored line: FI for position measurement $\mathcal{F}_c$. The FI for complete graph and cycle graph (circulant graphs) is null, and therefore is not shown. Because of the different ranges, values of QFI are referred to the left $y$-axis, and values of FI are referred to the right $y$-axis.}
    \label{fig:FIQFI_graphs}
\end{figure*}

\begin{figure*}[!htb]
    \centering
    \includegraphics[width=\textwidth]{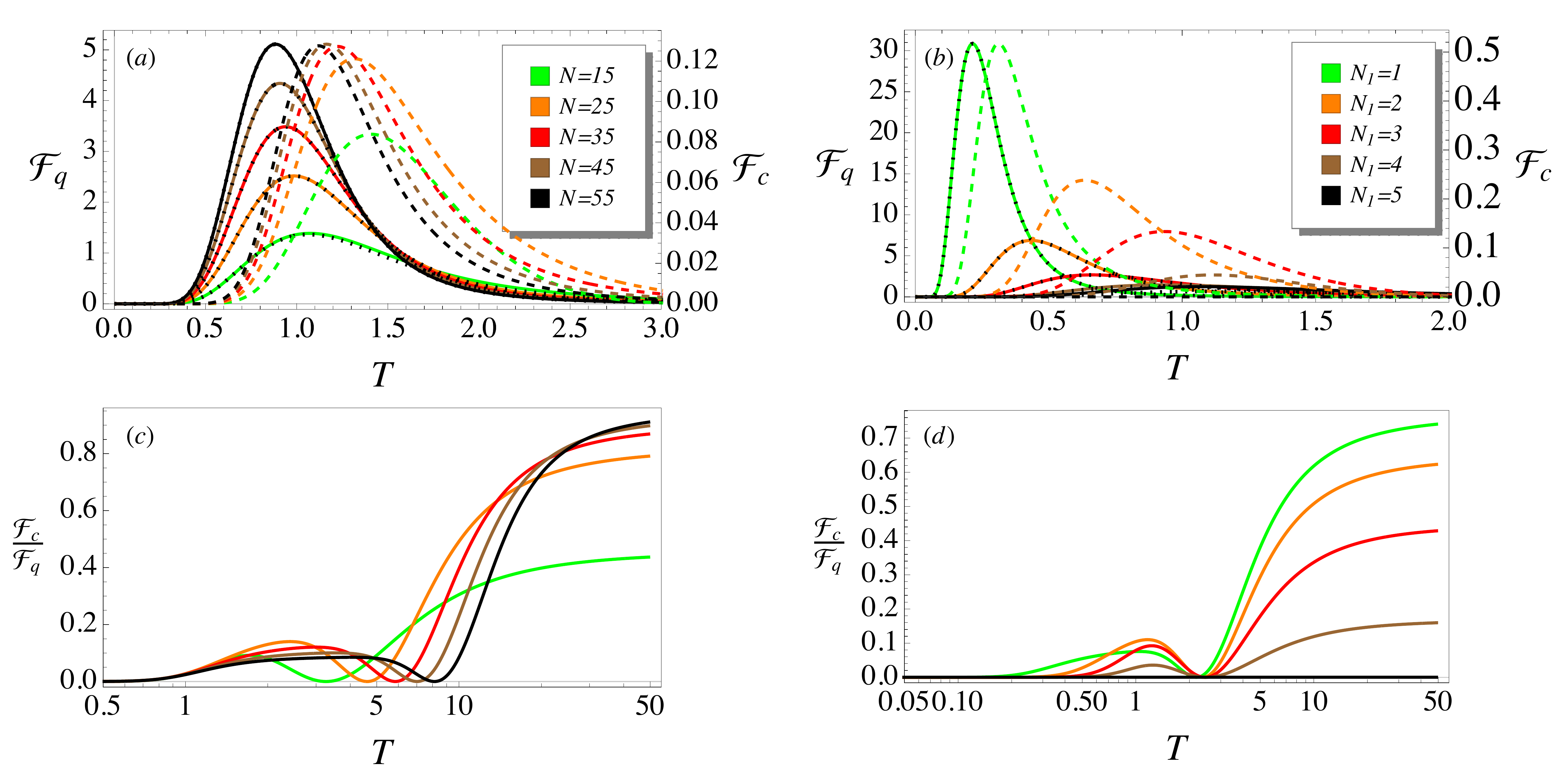}
     \caption{Results of the estimation problem of the temperature for the complete bipartite graph $K_{N_1,N_2}$ of order $N=N_1+N_2$. 
     Left-column plots: results for different $N$ at fixed $N_1=5$. Right-column plots: results for different $N_1$ at fixed $N=10$. Values of $N_1 > N/2$ are not considered because of the symmetry of the graph when exchanging the two partite sets and so $N_1$ and $N_2$.
     For $N_1=N_2=N/2$ the FI is identically null because the corresponding complete bipartite graph is circulant. Top-row plots: QFI $\mathcal{F}_q$ (solid colored line), QFI at low temperature $\mathcal{F}_q^{low}$ \eqref{eq:lowTqfi_approx} (dotted black line), and FI for position measurement $\mathcal{F}_c$ (dashed colored line). Because of the different ranges, values of QFI are referred to the left $y$-axis, and values of FI are referred to the right $y$-axis. Bottom-row plots: ratio $\mathcal{F}_c/\mathcal{F}_q$.}
    \label{fig:complete_bipartite_graph}
\end{figure*}

In this section, we address the study for some remarkable connected simple graphs and some lattice graphs by means of the previously found general results. To avoid repetitions, we recall that the ground state energy $E_0 =0$ is not degenerate for connected simple graphs, $g_0=1$, and the corresponding eigenstate is
\begin{equation}
    \vert e_0 \rangle = \frac{1}{\sqrt{N}}\sum_{k=0}^{N-1} \vert k \rangle\,.
\end{equation}
Results of QFI and FI for position measurement for graphs (see Fig. \ref{fig:graphs}) are shown in Fig. \ref{fig:FIQFI_graphs} and \ref{fig:complete_bipartite_graph}, for  lattices (see Fig. \ref{fig:lattices}) in Fig. \ref{fig:FIQFI_lattices}, and results of the ratio of FI and QFI for both graphs and lattices are summarized in Fig. \ref{fig:FIQFI_ratio}. The analytical results suitable for a comparison are reported in Table \ref{tab:Q_FI_comparison}.

\subsection{Complete graph}
A complete graph is a simple graph whose vertices are pairwise adjacent, i.e. each pair of distinct vertices is connected by a unique edge (see Fig. \ref{fig:graphs}(a)). The complete graph with $N$ vertices is denoted $K_N$, is $(N-1)$-regular, and has $M=N(N-1)/2$ edges. Its energy spectrum consists of two energy levels: the ground state and the second level $E_1=N$ with degeneracy $g_1=N-1$. The graph is circulant, thus the eigenvectors are given by \eqref{eq:eigenvector_circulant} and the FI for a position measurement is identically null.

In this case, the approximation for the low-temperature regime is actually exact and holds at all the temperatures, because the system has precisely two distinct energy levels. Hence, the QFI reads as
\begin{equation}
\mathcal{F}_q(T) = \frac{ N^2 (N-1) e^{-N/T}}{T^4[1+(N-1) e^{-N/T}]^2}\,.
\label{eq:QFI_complete}
\end{equation}
The algebraic connectivity $E_1=N$ and the degeneracy $g_1=N-1$ grow with the order $N$ of the graph. In Fig. \ref{fig:FIQFI_graphs}(a) we observe that maxima of QFI occur at higher temperatures as $N$ increases. According to Eq. \eqref{eq:transcendental} and Fig. \ref{fig:graph_sol_maxQFI}, we expect the maximum of QFI to occur at increasing values of $x_{max} = E_1/T_{max}$ as $g_1$ ($N$) increases. Hence, this means that $T_{max}$ increases less than linearly with $N$. For this reason the complete graph is not a good thermometer for low $T$. On the other hand, the complete graph saturates the upper bound in \eqref{eq:degreesquared}, since $M=N(N-1)/2$. It follows that in the high-temperature regime the complete graph is the optimal thermometer and, accordingly, the QFI is $\mathcal{F}^{high}_q(T) = (N-1)/T^4$.

\subsection{Cycle graph}
A cycle graph with $N\geq 3$ vertices (or $N$-cycle) is a simple graph whose vertices $\lbrace v_j\rbrace_{j=1,\ldots, N}$ can be (re)labeled such that its edges are $v_1v_2, v_2v_3,\ldots,v_{N-1}v_N$, and $v_{N}v_1$ (see Fig. \ref{fig:graphs}(b)). In other words, we may think of it as a one-dimensional lattice with $N$ sites and periodic boundary conditions. The cycle graph with $N$ vertices is denoted $C_N$, is $2$-regular, and has $M=N$ edges. Its energy spectrum is $E_k =2[1-\cos(2\pi k/N)]$, with $k=0,\ldots,N-1$. The lowest energy level is not degenerate, while the degeneracy of the highest energy level depends on the parity of $N$: no degeneracy for even $N$, $g_{N/2} =1$, but double degeneracy for odd $N$, $g_{(N+1)/2} = 2$. The remaining energy levels have degeneracy $2$. The cycle graph is circulant, thus the eigenvectors are \eqref{eq:eigenvector_circulant}, the same of those of the complete graph, and the FI for a position measurement is identically null.

The algebraic connectivity $E_1 =2[1-\cos(2\pi/N)]$ decreases as $N$ increases, while $g_1=2$ is constant. According to Eq. \eqref{eq:transcendental} and Fig. \ref{fig:graph_sol_maxQFI}, we expect the maximum of QFI to occur at the constant value of $x_{max} = E_1/T_{max}$ independently of $N$, because $g_1$ is constant. Since $E_1$ decreases as $N$ increases, then $T_{max}$ must also decrease to ensure $x_{max}$ constant. Indeed, the maxima of QFI occur at lower temperatures as $N$ increases, as shown in Fig. \ref{fig:FIQFI_graphs}(b). It follows that the larger $N$ the better the cycle graph behaves as a low-temperature probe. Instead, the cycle graph saturates the lower bound in \eqref{eq:degreesquared}, since $M=N$, and so the QFI at high temperatures is $\mathcal{F}^{high}_q(T)=2/T^4$.

\subsection{Complete Bipartite Graph}
\label{sec:CBG}
A graph $G$ is bipartite if the set of vertices $V(G)$ is the union of two disjoint independent sets $V_1$ and $V_2$, called partite sets of $G$, such that every edge of $G$ joins a vertex of $V_1$ and a vertex of $V_2$. A complete bipartite graph is a simple bipartite graph such that two vertices are adjacent if and only if they are in different partite sets, i.e. if every vertex of $V_1$ is adjacent to every vertex of $V_2$ (see Fig. \ref{fig:graphs}(c)). The complete bipartite graph having partite sets with $\vert V_1 \vert =N_1$ and $\vert V_2 \vert =N_2$ vertices is denoted $K_{N_1,N_2}$, has $M = N_1 N_2$ edges, and  the total number of vertices is $N=N_1+N_2$.
Without loss of generality we assume $N_1\leq N_2$. The energy spectrum is given by $E_1=N_1$, $E_2=N_2$, and $E_3=N_1+N_2$, with degeneracy $g_0=1$, $g_1=N_2-1$, $g_2=N_1-1$, and $g_3=1$, respectively. The corresponding eigenvectors are
\begin{widetext}
\begin{align}
    \ket{e^n_1} = \frac{1}{\sqrt{n(n+1)}}\left(\sum_{k=N_1}^{N_1-1+n}\ket{k}-n\ket{N_1+n}\right) \,,&\quad 
    \ket{e^m_2} = \frac{1}{\sqrt{m(m+1)}}\left(\sum_{k=0}^{m-1}\ket{k}-m\ket{m}\right) \,,\nonumber\\
    \ket{e_3} = \frac{1}{\sqrt{N}}\left(\sqrt{\frac{N_2}{N_1}}\sum_{k=0}^{N_1-1}\ket{k}-\sqrt{\frac{N_1}{N_2}}\sum_{k=N_1}^{N-1}\ket{k}\right)\,,&
\end{align}
\end{widetext}
where $n=1,\ldots,N_2-1$ and $m=1,\ldots,N_1-1$.
\par
Note that for $N_1=N_2=N/2$ the complete bipartite graph is circulant \cite{WolframCBG} and the spectrum reduces to $E_0$, $E_1 = N/2$, and $E_2 =N$, with degeneracy, respectively, $g_0=1$, $g_1=N-2$, and $g_2=1$. Instead, for $N_1=1$ and $N_2=N-1$ we obtain the star graph $S_N$, whose spectrum reduces to $E_0$, $E_1 = 1$, and $E_2 =N$, with degeneracy, respectively, $g_0=1$, $g_1=N-2$, and $g_2=1$.
\par
Regarding the low-temperature regime, the algebraic connectivity is $E_1=N_1$ while $g_1 =N_2-1$. 
The complete bipartite graph is completely defined only by the total number of vertices $N$, so we discuss where the maximum of the QFI occur according to Eq. \eqref{eq:transcendental} and Fig. \ref{fig:graph_sol_maxQFI} first for a given value $N_1$, and then for a given value of $N=N_1+N_2$.

For $N_1$ fixed, we expect the maximum of QFI to occur at increasing values of $x_{max} = E_1/T_{max}$ as $N$ increases, because $N_2$ and thus $g_1$ increase. Since $E_1$ is constant, then $T_{max}$ must decrease to ensure that $x_{max}$ increases. Indeed, for a given $N_1$, the maxima of QFI occur at lower temperatures as $N$ increases, as shown in Fig. \ref{fig:complete_bipartite_graph}(a). In particular, this is also the case of the star graph $S_N$, because it is $K_{1,N-1}$, even if such behavior is less evident in  Fig. \ref{fig:FIQFI_graphs}(c).

For $N$ fixed, we expect the maximum of QFI to occur at decreasing values of $x_{max} = E_1/T_{max}$ as $N_1$ increases, because $N_2$ and thus $g_1$ decrease. Since $E_1$ increases as $N_1$ increases, then $T_{max}$ must increase more than $N_1$ to ensure that $x_{max}$ decreases. Indeed, for a given $N$, the maxima of QFI occur at higher temperatures as $N_1$ increases, as shown in Fig. \ref{fig:complete_bipartite_graph}(b). This means that, at fixed $N$, we can tune the temperature at which the QFI is maximum just by varying the number of of vertices in the two partite sets. From Fig. \ref{fig:complete_bipartite_graph}(b) we observe that the highest maximum of QFI is provided by the star graph $S_N$, whose algebraic connectivity $E_1=1$ is constant and minimum, while the lowest maximum of QFI is provided by $K_{N/2,N/2}$, i.e. for $N_1=N_2$, whose algebraic connectivity $E_1=N/2$ is the largest among all the complete bipartite graphs. 
\par
In the high-temperature regime, since $\sum_k d_k^2=N_1 N_2(N_1+N_2)$ and $M=N_1N_2$, the QFI is  
\begin{equation}
\mathcal{F}^{high}_q(T)=\frac{N_1N_2[(N_1-N_2)^2+2(N_1+N_2)]}{T^4(N_1+N_2)^2}\,.
\end{equation}
Notice that for $N_1=N_2=N/2$, the complete bipartite graph is $N/2$-regular, and saturates the lower bound in \eqref{eq:degreesquared}, since $M=N^2/4$, and so the QFI at high temperatures is $\mathcal{F}^{high}_q(T)=N/(2T^4)$.
\par
The asymptotic behavior of the ratio $\mathcal{F}_c(T)/\mathcal{F}_q(T)$ at high temperature \eqref{eq:limit_ratio_FIQFI} is characterized by $\lambda_{N_1+N_2,N_1N_2} = 2(N_1+N_2)/(N_2-N_1)^2$. Depending on the number of vertices in the two subsets, results differ. When $N_1 = N_2$, the difference $N_2-N_1$ is null, the complete bipartite graph is circulant and so the FI is identically null, for any $T$. Instead, the difference $N_2-N_1$ is maximum for the star graph $S_N$. This results in $\lambda_{N,N}= 2N/(N-2)^2$: hence, $\lambda_{N,N} \to 0$ for large $N$ and, accordingly, the FI approaches the QFI in the limit of high temperatures. Actually, since for the star graph $\sum_k d_k^2=N(N-1)$, the QFI in the high-temperature regime has the same asymptotic behavior of the complete graph, i.e. $\mathcal{F}_q^{high}= (N-1)/T^{4}+O(1/(NT^4))$. 
\par
In this section we have approximated the QFI for the complete bipartite graph under the assumptions of low or high temperature. The exact analytical expression of the QFI is reported in Appendix \ref{app:QFIcbg}.

\subsection{Path graph}
A path graph with $N$ vertices is a simple graph whose vertices $\lbrace v_j\rbrace_{j=1,\ldots, N}$ can be (re)labeled such that its edges are $v_1v_2, v_2v_3,\ldots,v_{N-1}v_N$ (see Fig. \ref{fig:graphs}(d)). In other words, we may think of it as a one-dimensional lattice with $N$ sites and open boundary conditions. The path graph with $N$ vertices is denoted $P_N$, and has $M=N-1$ edges. Its nondegenerate energy spectrum is $E_k=2[1-\cos(\pi k/N)]$, with $k=0,\ldots,N-1$, and the corresponding eigenvectors are
\begin{equation}
    \ket{e_k} = \sum_{j=0}^{N-1}\cos(\frac{\pi k}{2N}(2j-1))\ket{j}. 
\end{equation} 
The energy spectrum is similar to that of the cycle, and this is reflected in its thermometric behavior. Indeed, the algebraic connectivity $E_1=2[1-\cos(\pi/N)]$ decreases as $N$ increases, while $g_1=1$ is constant. Hence, as for the cycle graph, the maximum of the QFI occurs at lower temperature as $N$ increases, as shown in Fig. \ref{fig:FIQFI_graphs}(d). Further, the similarity extends also in the high-temperature  regime, where, due to $\sum_{k}d_k^2=2(2N-3)$ and $M=N-1$, we have that $\mathcal{F}^{high}_q(T) = 2/T^{4} + O(1/(N^2T^4))$, which is asymptotically equivalent to that of the cycle.
\par
Nevertheless, there is a difference between the cycle and the path, and this is due to the different boundary conditions of the two graphs. In the first, the periodic boundary conditions ensure that the cycle graph is a circulant graph, and consequently the FI for the position measurement is null. Instead, in the second, the open boundary conditions lead to a non-null FI for the position measurement. The asymptotic behavior of the ratio $\mathcal{F}_c(T)/\mathcal{F}_q(T)$ at high temperature \eqref{eq:limit_ratio_FIQFI} is characterized by $\lambda_{N,N-1} = N(N-1)/(N-2)$, which is monotonically increasing with the order of the graph. Thus, in the limit of high temperature the FI is very small compared to QFI. 

\begin{figure}[!htb]
\centering
\includegraphics[width = 0.45\textwidth]{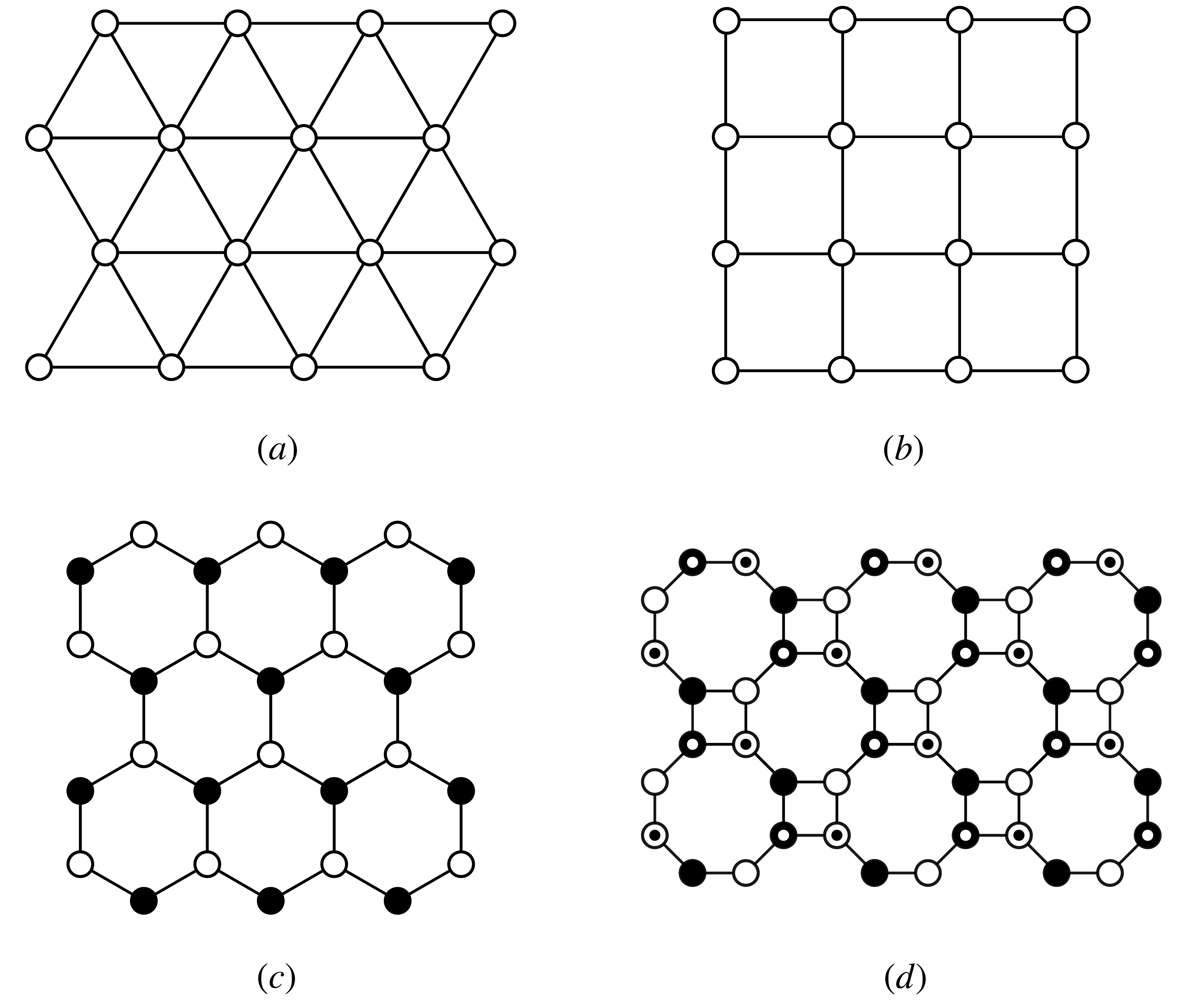}
\caption{Two-dimensional lattices considered in the present work: (a) triangular, (b) square, (c) honeycomb, and (d) truncated square lattice. Equivalent vertices are equally represented.}
\label{fig:lattices}
\end{figure}

\begin{figure*}[!htb]
    \centering
    \includegraphics[width=\textwidth]{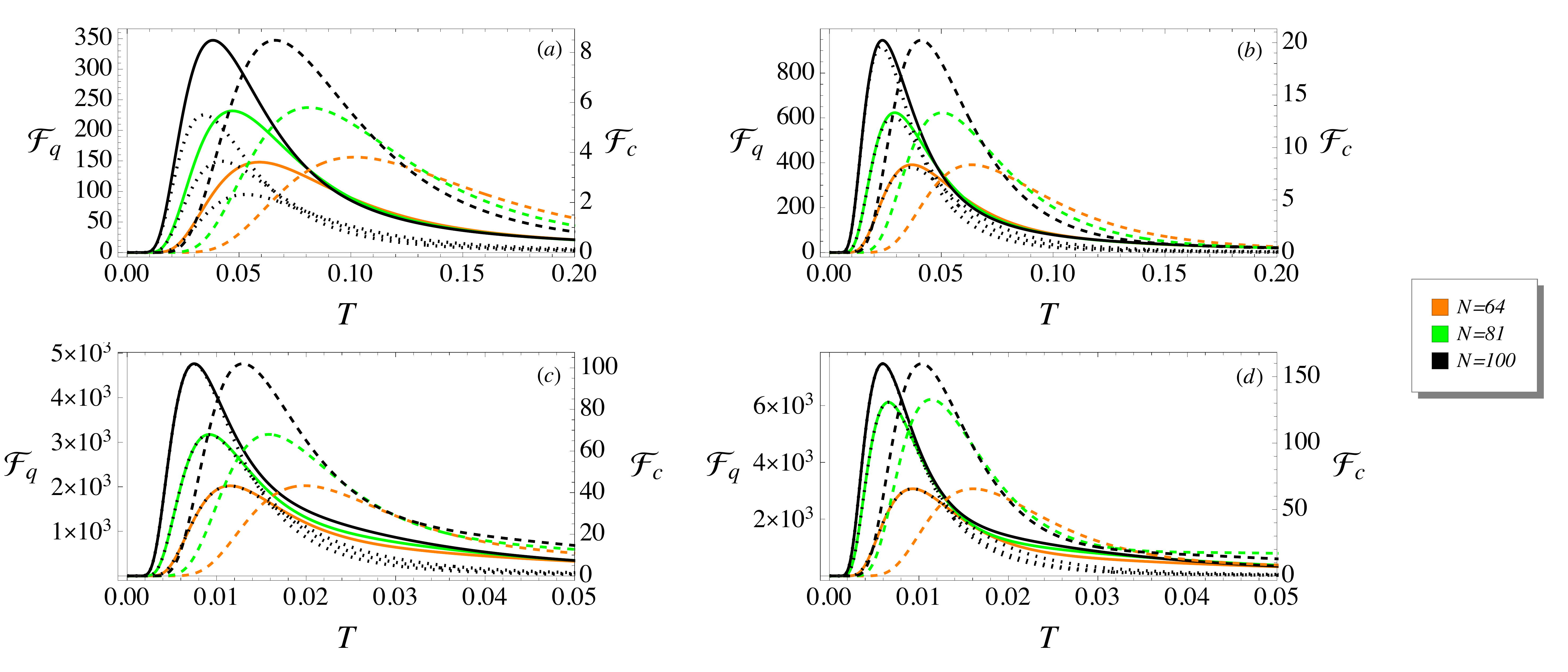}
    \caption{QFI and FI for position measurement for different $\sqrt{N} \times \sqrt{N}$ lattices with open boundary conditions (OBC): (a) Triangular lattice, (b) square lattice, (c) honeycomb lattice, and (d) truncated square lattice. Solid colored line: QFI $\mathcal{F}_q$. Dotted black line: QFI at low temperature $\mathcal{F}_q^{low}$ \eqref{eq:lowTqfi_approx}. Dashed colored line: FI for position measurement $\mathcal{F}_c$. Because of the different ranges, values of QFI are referred to the left $y$-axis, and values of FI are referred to the right $y$-axis.}
    \label{fig:FIQFI_lattices}
\end{figure*}

\subsection{Lattices}
In this section we address the thermometry on some two-dimensional lattices. There are three regular tessellations composed of regular polygons symmetrically tiling the Euclidean plane: equilateral triangles, squares, and regular hexagons [Figs. \ref{fig:lattices}(a)-(c)]. In addition to these we also consider the truncated square lattice in Fig. \ref{fig:lattices}(d). Triangular and square lattices are Bravais lattices, while honeycomb and truncated square lattice are not. This difference is reflected in the spreading of CTQWs, which is ballistic on Bravais lattices and subballistic on non-Bravais lattices \cite{razzoli2020continuous}. A generic vertex in the triangular lattice has degree 6, in the square lattice has degree 4, and both in the honeycomb and in the truncated square lattice has degree 3. We consider the lattices either with open boundary conditions (OBC) or with periodic boundary conditions (PBC). Notice that the lattices  with PBC are regular, while the lattices with OBC are not, because the vertices at the boundaries have a lower degree than the vertices within the lattice.

\begin{figure*}[!htb]
    \centering
    \includegraphics[width=\textwidth]{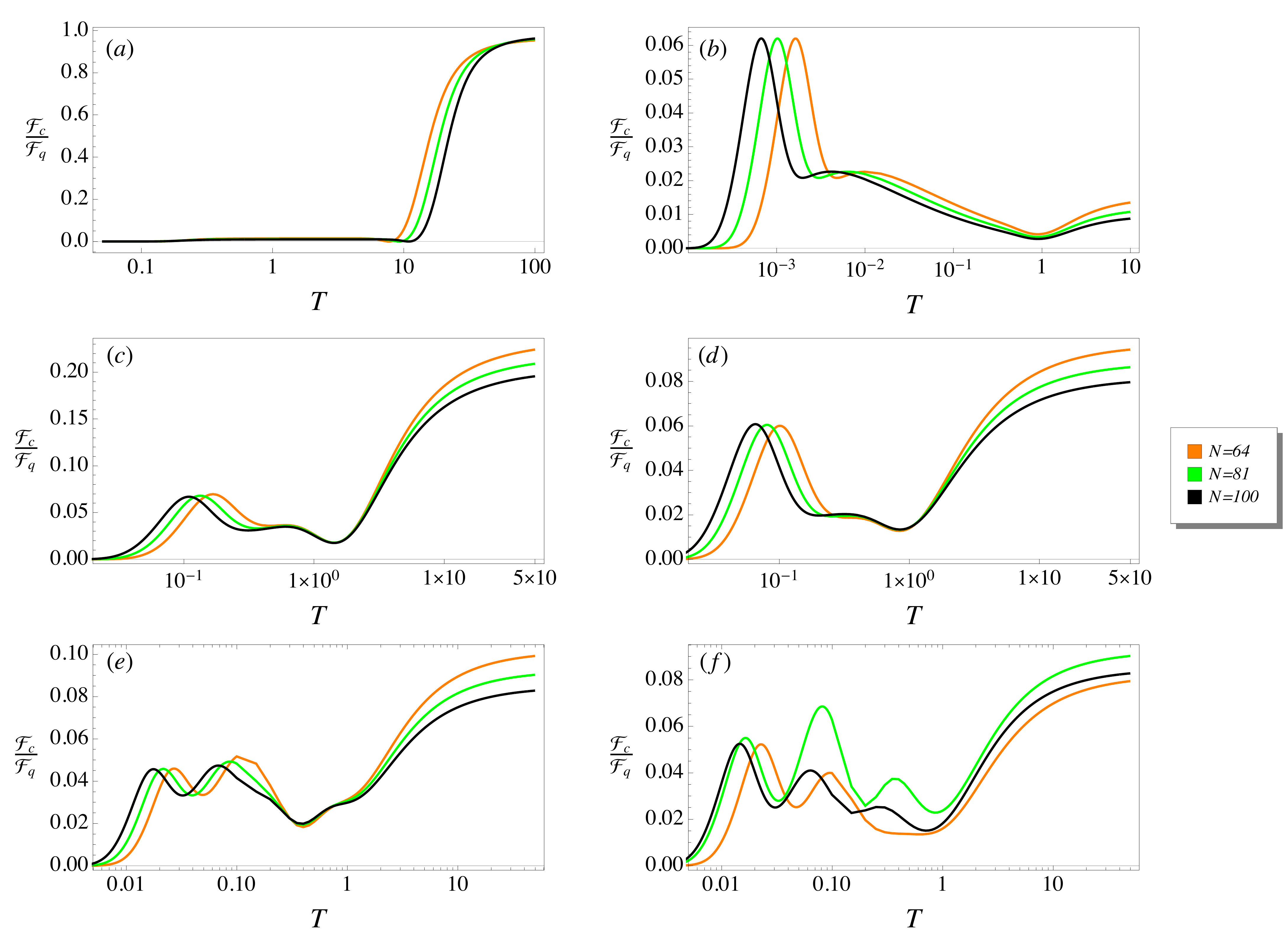}
    \caption{Ratio $\mathcal{F}_c/\mathcal{F}_q$ of FI and QFI for the graphs of order $N$ and the $\sqrt{N} \times \sqrt{N}$ lattices providing non-null FI. (a) Star graph, (b) path graph, (c) triangular lattice  (OBCs), (d) square lattice (OBCs), (e) honeycomb lattice (OBCs), and (f) truncated square (OBCs). Note the logarithmic scale of the temperature axis.}
    \label{fig:FIQFI_ratio}
\end{figure*}

Numerical results of QFI and FI for the lattices with OBC are shown in Fig. \ref{fig:FIQFI_lattices}. We observe that the maximum of the QFI gets sharper and higher, and shifts to lower temperatures as the size of the lattice, i.e., the number of vertices, increases. A similar behavior occurs as the degree of the vertex of the lattice decreases: the maximum of the QFI for honeycomb and truncated square lattices is sharper and higher, and at lower temperature than the peak of the QFI for the triangular lattice. The predicted behavior of the QFI at low temperature \eqref{eq:lowTqfi_approx} is a good approximation for honeycomb and truncated square lattices, because it fits the maximum of the QFI, its height and position. For the square it is fairly good approximation, but for the triangular lattices it fits only the QFI at the temperatures closer to zero. The FI of position measurement is a couple of orders of magnitude lower than the QFI [see the ratio $\mathcal{F}_c(T)/\mathcal{F}_q(T)$ in Fig. \ref{fig:FIQFI_ratio}], and its maximum is at higher temperature than the maximum of the QFI. 

For lattices with PBCs the behavior of the QFI is qualitatively the same as regards the goodness of the lower-temperature approximation \eqref{eq:lowTqfi_approx} and the dependence of the QFI on the size of the lattice and the degree of the vertices. However, the maxima of QFI for lattices with PBCs are lower and occur at higher temperature than the maxima of QFI for lattices with OBCs. Remarkably, the FI for these lattices with PBCs is identically null.

Some analytical results can be obtained for the square lattice, both with OBCs and with PBCs. Indeed, the $m \times n$ square lattice with OBCs is actually a grid graph and is the Cartesian product of two path graphs, $G_{m,n} = P_m \square P_n$ \cite{WolframGridGraph}. Instead, the $m \times n$ square lattice with PBCs is actually the torus grid graph and is the Cartesian product of two cycle graphs, $T_{m,n} = C_m \square C_n$ \cite{WolframTorusGridGraph}. For the Cartesian product $G_1 \square G_2$ of two graphs $G_1$ and $G_2$ we can easily obtain the QFI and FI as follows (proof in Appendix \ref{app:qfifi_cartprod}):
\begin{align}
\mathcal{F}_q(G_1 \square G_2 \vert T) &= \mathcal{F}_q(G_1 \vert T) + \mathcal{F}_q (G_2 \vert T)\,,\label{eq:QFIcartprod}\\
\mathcal{F}_c(G_1 \square G_2 \vert T) &= \mathcal{F}_c(G_1 \vert T) + \mathcal{F}_c(G_2 \vert T)\,.\label{eq:FIcartprod}
\end{align}
Thus, since the FI of position measurement for the cycle graph is identically null, this result analytically proves the null FI for the square lattice with PBCs.

\begin{table*}[tb]
    \centering
    \caption{QFI $\mathcal{F}_q^{low}$ \eqref{eq:lowTqfi_approx} in the low-temperature regime and QFI $\mathcal{F}_q^{high}$ \eqref{eq:QFI_highT}, FI $\mathcal{F}_c^{high}$ \eqref{eq:FI_highT}, and their ratio in the high-temperature regime for the graphs considered in the present work: complete graph $K_N$, cycle graph $C_N$, complete bipartite $K_{N_1,N_2}$, star graph $S_N$, and path graph $P_N$. Analytical results are also available for the $\sqrt{N} \times \sqrt{N}$ square lattice with OBC (grid graph $G_{\sqrt{N},\sqrt{N}}$) and with PBC (torus grid graph $T_{\sqrt{N},\sqrt{N}}$), since grid graph and torus grid graph are the Cartesian product of two path graphs and two cycle graphs respectively (see Appendix \ref{app:qfifi_cartprod}). To have a fair comparison in terms of the total number of vertices $N$, in the table we report the result for $\sqrt{N}\times\sqrt{N}$ square lattices, and for the complete bipartite graph $K_{N_1,N_2}$ we write results as a function of $N=N_1+N_2$ and $\Delta = N_2-N_1$ ($N_2 \geq N_1$ as assumed in the paper), except for the QFI in the low-temperature regime. The FI $\mathcal{F}_c^{low}$ in the low-temperature regime is not reported, because an expression suitable for a comparison is not available (see Eq. \eqref{eq:lowTfi}). Both QFI and FI in the high-temperature regime depend on the temperature as $T^{-4}$, thus we report their values multiplied by $T^4$ to focus on the factor which depends on the topology of the graph. The same criterion is adopted for the QFI in the low-temperature regime for consistency. Numerical results show that graphs with low degree, e.g. $C_N$ and $P_N$, exhibit the highest maxima of the QFI at low temperatures. Conversely, at high temperatures and at fixed $N$, the maximum QFI is obtained with the complete and the star graph, whose QFI scales linearly with the order $N$. Indeed, in the limit of $N\to \infty$, the QFI of $P_N$ approaches that of $C_N$, as well as the QFI of $G_{\sqrt{N},\sqrt{N}}$ approaches that of $T_{\sqrt{N},\sqrt{N}}$.}
    \label{tab:Q_FI_comparison}
    \begin{ruledtabular}
        \begin{tabular}{ccccc}
        & Low-temperature  & \multicolumn{3}{c}{High-temperature}\\ \cline{2-2}\cline{3-5}
        Graph & $T^4\mathcal{F}_q^{low}$ &  $T^4\mathcal{F}_q^{high}$ & $T^4\mathcal{F}_c^{high}$ & $\mathcal{F}_c^{high}/\mathcal{F}_q^{high}$\\\hline
        &\\
        $K_N$ & $\frac{ N^2  (N-1) \exp(-N/T)}{[1 + (N-1)\exp(-N/T)]^2}$ & $N-1$ & $0$ & $0$\\[6pt]
        $C_N$& $\frac{32 \exp[-4 \sin^2(\pi/N)/T]
  \sin^4(\pi /N)}{\left(1+2\exp[-4 \sin^2(\pi/N)/T]\right)^2}$ & $2$ & $0$ & $0$\\[6pt]
         $K_{N_1,N_2}$& $\frac{\exp(-N_1/T) N_1^2 (N_2-1)}{[1+(N_2-1)\exp(-N_1/T)]^2}$& $\frac{(N^2-\Delta^2)(\Delta^2+2N)}{4N^2}$ & $\frac{(N^2-\Delta^2)\Delta^2}{4N^2}$ &  $\frac{1}{1+2N/\Delta^2}$\\[6pt]
       $S_N$& $\frac{(N-2)\exp(-1/T)}{[1 + (N-2)\exp(-1/T)]^2}$ & $\frac{(N-1) [N(N-2)+4]}{N^2}$ & $\frac{(N-1)(N-2)^2}{N^2}$ & $\frac{(N-2)^2}{N(N-2)+4}$\\[6pt]
       $P_N$ & $\frac{16 \exp[-4 \sin^2(\pi/2N)/T]
  \sin^4(\pi /2N)}{\left(1+\exp[-4 \sin^2(\pi/2N)/T]\right)^2}$ & $\frac{2(N^2-2)}{N^2}$ & $\frac{2(N-2)}{N^2}$ & $\frac{N-2}{N^2-2}$\\[6pt]
  $G_{\sqrt{N},\sqrt{N}}$ & $\frac{32 \exp[-4 \sin^2(\pi/2\sqrt{N})/T]
  \sin^4(\pi /2\sqrt{N})}{\left(1+2\exp[-4 \sin^2(\pi/2\sqrt{N})/T]\right)^2}$ & $\frac{4(N-2)}{N}$ & $\frac{4(\sqrt{N}-2)}{N}$ & $\frac{\sqrt{N}-2}{N-2}$\\[6pt]
  $T_{\sqrt{N},\sqrt{N}}$& $\frac{64 \exp[-4 \sin^2(\pi/\sqrt{N})/T]
  \sin^4(\pi /\sqrt{N})}{\left(1+4\exp[-4 \sin^2(\pi/\sqrt{N})/T]\right)^2}$ & $4$ & $0$ & $0$\\
        \end{tabular}
    \end{ruledtabular}    
\end{table*}

\section{Role of coherence}
Temperature is a classical parameter, i.e. any change in the temperature modifies the eigenvalues of the Gibbs state but not the eigenvectors, which coincide with the eigenvectors of the Hamiltonian at any temperature. 
As a consequence, one may wonder whether  quantumness is playing any role in our analysis, which also does not rely upon quantum effects as entanglement.  Despite the above arguments, the quantum nature of the systems under investigation indeed plays a role in determining topological effects in thermometry. In fact, 
thermal states \eqref{eq:equilstate} are diagonal in the Hamiltonian basis, but show quantum coherence in the position basis, which itself is the reference {\em classical basis} when looking at topological effects in graphs. In turn, as we will see in the following, the peak of the QFI occurs in the interval of temperatures over which the coherence starts to decrease.
\par
In order to quantitatively assess the role of coherence, let us consider the $l_1$ norm of coherence \cite{baumgratz2014quantifying}
\begin{equation}
\mathcal{C}(\rho)=\sum_{\substack{j,k=0,\\j\neq k}}^{N-1} \abs{\rho_{j,k}}
\label{eq:coherence_def}
\end{equation}
as a measure of quantum coherence of a state $\rho$. For convenience, we normalize this measure to its maximum value $\mathcal{C}(\rho_N)=N-1$, thus defining $C(\rho):=\mathcal{C}(\rho)/(N-1)$. At $T=0$, the system is at thermal equilibrium in its ground state and since the Hamiltonian of the system is the Laplacian of a simple graph, the ground state is the maximally coherent state $\vert \psi_N \rangle = \sum_{j=1}^N \vert j\rangle/\sqrt{N}$. The normalized coherence is thus equal to one.

\begin{figure}[h]
    \centering
    \includegraphics[width=0.48\textwidth]{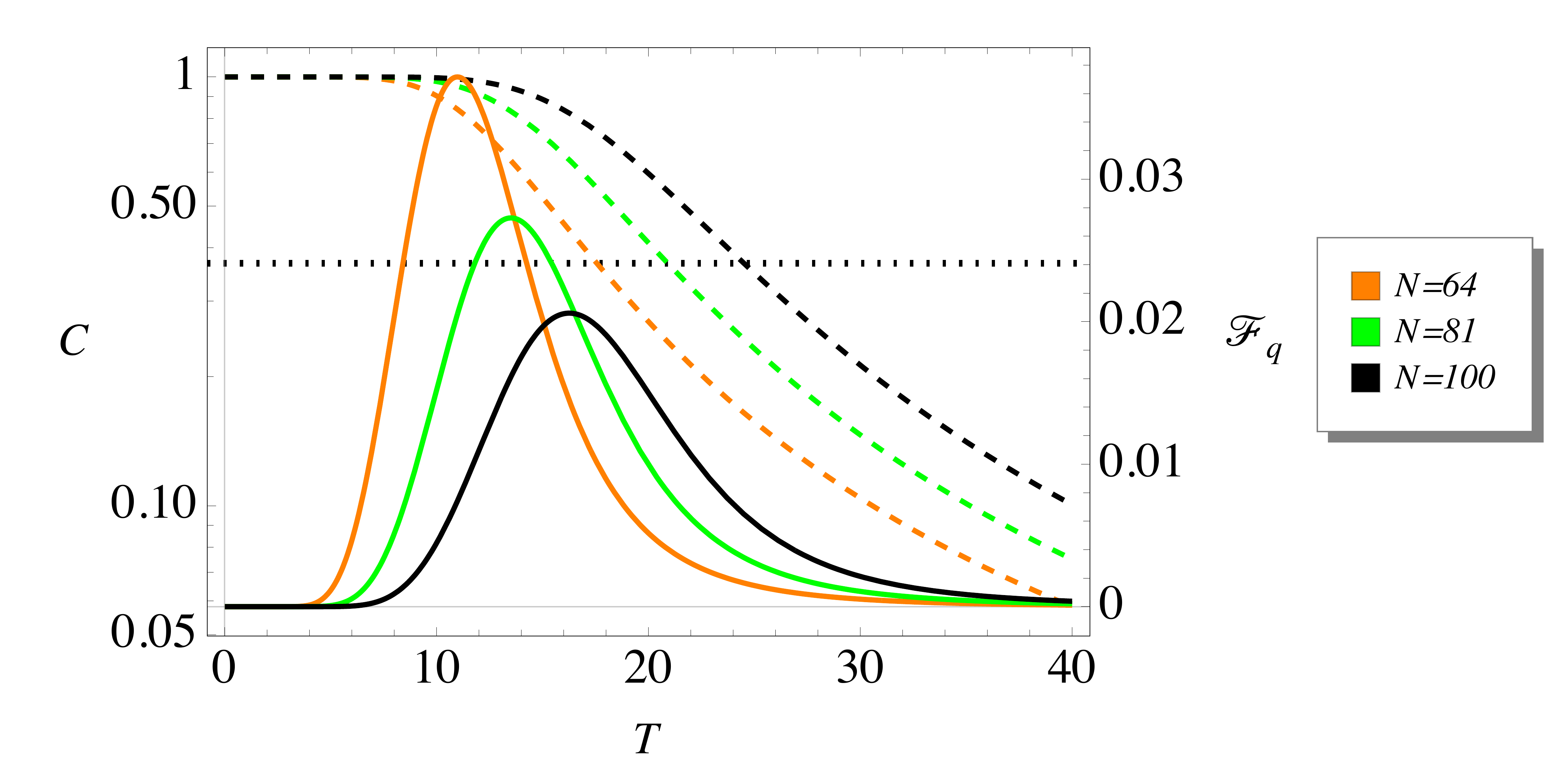}
    \caption{QFI $\mathcal{F}_q$ (solid line) and normalized coherence $C$ (dashed line) of a Gibbs state $\rho_T$ as a function of $T$ for a complete graph of order $N$. The black horizontal dotted line represents the constant value $1/e$.}
    \label{fig:cohQFI}
\end{figure}

As far as the temperature is very low, the ground state is robust, the coherence remains close to one, and the QFI is small, i.e. the robustness of the ground state prevents the system to effectively monitor any change in temperature. On the other hand, when temperature increases, thermal effects becomes more relevant, coherence decreases, and the QFI increases. In other words, it is the fragility of quantum coherence which makes the system a good sensor for temperature (a common feature in the field of {\em quantum probing}). For higher temperatures, the Gibbs state approaches a {\em flat} mixture, almost independent of temperature, and both the coherence and the QFI vanish. In order to illustrate the argument, let us consider the case of complete graphs, for which we have analytic expressions for the QFI, see Eq. \eqref{eq:QFI_complete}, and for the normalized coherence
\begin{equation}\label{qq}
    C(\rho_T) = \frac{\vert 1-e^{-N/T}\vert}{1+(N-1)e^{-N/T}}\,.
\end{equation}
As it is apparent from Fig. \ref{fig:cohQFI}, where we show the two quantities, the peak of QFI indeed occurs in the interval of temperatures over which the coherence is reduced by a factor $1/e$ (we have numerically observed analogous behavior also for the other graphs). Upon comparison of Eq.  \eqref{eq:QFI_complete} with Eq. \eqref{qq} we may also write 
\begin{align}
\frac{T^4\, \mathcal{F}_q (T)}{N-1} = \Big[1-C(\rho_T)\Big]\Big[1+(N-1) 
C(\rho_T)\Big]\,.
\end{align}

\section{Conclusion}
\label{sec:conclusion}
We have addressed the role of topology in determining the precision of thermometers. The key idea is to use a 
finite system as a probe for estimating the temperature $T$ of an external environment. 
The probe is regarded as a connected set of subunits and may be ultimately modeled 
as a quantum walker moving continuously in time on a graph. In particular, we have 
considered equilibrium thermometry, and evaluated the quantum Fisher information of 
Gibbs states. Since the Hamiltonian of a quantum walker corresponds to the Laplacian 
matrix of the graph, the topology is inherently taken into account.
We have considered some paradigmatic graphs and two-dimensional lattices, evaluated the Fisher information (FI) for a position measurement and compared it with the quantum Fisher information (QFI, energy measurement), providing analytical and numerical results. In particular, we have focused on the low- and the high-temperature regimes, which we have investigated by means of analytic approximations which allow us to have a better understanding of the behavior of the system.
\par
We have proved, by numerical and analytical means, that the maximum of the QFI and the corresponding optimal temperature depend on the two topological parameters of the graph: the algebraic connectivity and the degeneracy of the first energy level. In our system, the algebraic connectivity also represents the energy gap between the first excited energy level and the ground state, and the smaller is the algebraic connectivity, the higher is the maximum of the QFI. These results are supported by a number of examples. In particular, graphs and lattices whose vertices have low degree, e.g. path and cycle graphs, as well as honeycomb and truncated square lattices, show the highest maxima of QFI. We also notice that the maximum of the QFI and the corresponding optimal $T$ decrease as $N$ increase in the complete graph, while in all the other cases we have the opposite behavior.

At intermediate temperatures, the analytical approximation we have at low temperatures is no longer valid, as shown by the discrepancy between the dotted lines (analytical approximation) and the solid lines (exact results) in Figs. \ref{fig:FIQFI_graphs}--\ref{fig:complete_bipartite_graph}, and \ref{fig:FIQFI_lattices}). However, the low-temperature approximation captures quite well the maximum of the QFI, after which the QFI decreases, tending to zero, as the temperature increases. This behavior is confirmed by the exact analytical expressions of the QFI we have for the complete graph, Eq. \eqref{eq:QFI_complete}, and the complete bipartite graph (see Appendix \ref{app:QFIcbg}), and we also have numerical evidence of it for the other graphs and lattices. Hence, no relevant structures of the QFI are expected at intermediate temperatures.
\par
At high temperatures the QFI is of order $O(T^{-4})$, with a proportionality constant which depends on the topology of the graph. In this regime, the maximum QFI is attained by the complete graph, which is the simple graph that, at given number of vertices, has the highest number of edges. A remarkable thermometer is also obtained considering the complete bipartite graph. Despite its low QFI (if compared with the cycle and path graphs) it is possible to tune the position of the maximum of QFI just by varying the number of vertices in the two partite sets of the graph keeping fixed their sum. 
\par
Finally, we have discussed the role of coherence (in the position basis) in determining the precision. Our results provides some general indications on the role of topology in using quantum probes for thermometry, and provide new insights in the thermometry of finite-size quantum systems at equilibrium, at least for the class of systems where the Hamiltonian is in the form of a Laplacian matrix. 
In particular, our results suggest that quantum probes are particularly efficient in the low-temperatures regime, where the QFI reaches its maximum. They also pave the way to investigate the role of topology in out-of-equilibrium thermometry.

\acknowledgments
P.B. and M.G.A.P. are members of INdAM-GNFM.

\appendix
\section{Sum of the Laplacian eigenvalues and sum of their square}
\label{app:qfihighT}
First, we focus on the sum of the Laplacian eigenvalues $E_k$
\begin{equation}
	\sum_{k=0}^{N-1} E_k = \Tr[L] =  \Tr[D] = \sum_{k=0}^{N-1} d_k = 2 M,
\end{equation}
where the last equality was first proved by Euler and it is known as the degree sum formula or the handshaking lemma \cite{euler1741solutio,aldous2003graphs}. 
\par
Next, we write the sum of the $E_k^2$ as
\begin{align}
	\sum_{k=0}^{N-1} E_k^2 & = \Tr [L^2] \nonumber\\
	& = \Tr[D^2]-\Tr[A D]-\Tr[D A] + \Tr[A^2]\,.
\end{align}
Using the definition of degree and adjacency matrices, we see that
\begin{align}
	(D A)_{k,j} &=
	\begin{cases}
		0 & \text{if $k=j$,} \\
		d_k A_{k,j} & \text{otherwise,}
	\end{cases} \\
	(A D)_{k,j} &= 
	\begin{cases}
		0 & \text{if $k=j$,} \\
		A_{k,j}d_{j} & \text{otherwise,}
	\end{cases}	\\
	(A^2)_{k,j} &= \sum_{m=0}^{N-1} A_{k,m} A_{m,j}\,, 
\end{align}
and clearly $\Tr[DA]=\Tr[AD]=0$, whereas
\begin{align}
	\Tr[A^2] & = \sum_{k=0}^{N-1} (A^2)_{k,k} = \sum_{k=0}^{N-1} \sum_{m=0}^{N-1}  A_{k,m}A_{m,k} \nonumber \\
	& = \sum_{k=0}^{N-1} \sum_{m=0}^{N-1}A_{k,m} = \sum_{k=0}^{N-1} d_k = 2 M\,,
\end{align}
since the adjacency matrix is symmetric, $A_{k,m}=A_{m,k}$, and for simple graphs $A_{k,m} \in \lbrace 0,1 \rbrace$, thus $A_{k,m}^2=A_{k,m}$.

We also notice that this result is somehow related to the well-known fact that $(A^2)_{k,j}$ is the number of walk of length $2$ connecting the vertexes $k$ and $j$.  Eventually we obtain
\begin{equation}
	\sum_{k=0}^{N-1} E_k^2 = \sum_{k=0}^{N-1} d_{k}^2 + 2M\,.
\end{equation}

\section{Fisher Information for a position measurement}
\label{app:FIposition}
Let us consider the position measurement, whose POVM is given by $\{\vert j\rangle \langle j \vert \}$. Given an equilibrium state $\rho_T$ at temperature $T$, the probability distribution of the outcomes is given by the Born rule
\begin{align}
  p(j\vert T) &= \Tr\left[\rho_T \vert j \rangle \langle j \vert\right] = \sum_{k=0}^{N-1}  \frac{e^{-E_k/T}}{Z} \vert \langle j \vert e_{k} \rangle\vert^2,
\end{align}
and the FI by definition is \eqref{eq:FI_pos_def}.
From classical thermodynamics we recall that
\begin{gather}
\partial_T Z = \frac{Z \langle \hat{H} \rangle }{T^2},
\end{gather} 
and the first derivative of the probability distribution is
\begin{align}
  \partial_T p(j\vert T) & = \sum_{k=0}^{N-1}  \partial_T \left(\frac{e^{-E_k/T}}{Z}\right) \vert \langle j \vert e_{k} \rangle\vert^2  \nonumber \\
  &= \frac{1}{T^2} \sum_{k=0}^{N-1} e^{-E_k/T}\left(\frac{E_k - \langle \hat{H} \rangle}{Z}\right) \vert \langle j \vert e_{k} \rangle\vert^2 \nonumber \\ 
  & = \frac{1}{T^2} \left(\langle \hat{H}\rho_T \rangle_j - \langle \hat{H} \rangle p(j\vert T)\right)\,,
\end{align}
where $\langle \hat{H}\rho_T\rangle_j$ is given in Eq. \eqref{eq:EnergyWeighted}. From this result, the FI simplifies as
\begin{align}
\mathcal{F}_c(T) & =\frac{1}{T^4} \sum_{j=0}^{N-1}\frac{1}{p(j\vert T)}\Big( \langle \hat{H} \rho_T \rangle_j^2 + \langle \hat{H} \rangle^2 p(j\vert T)^2  \nonumber\\
& \quad -  2 \langle \hat{H}\rho_T \rangle_j \langle \hat{H} \rangle p(j\vert T) \Big) \nonumber \\ 
& = \frac{1}{T^4}  \sum_{j=0}^{N-1} \frac{\langle \hat{H}\rho_T \rangle_j^2}{p(j\vert T)} + \frac{1}{T^4}\langle \hat{H} \rangle^2 \sum_{j=0}^{N-1}p(j\vert T) \nonumber\\
& \quad - \frac{2}{T^4}\langle \hat{H} \rangle  \sum_{j=0}^{N-1}\langle \hat{H}\rho_T \rangle_j\,.
\end{align}
Since $ \sum_{j=0}^{N-1} \vert \langle j \vert e_k\rangle \vert^2 = 1$, we observe that
\begin{gather}
\sum_{j=0}^{N-1} \langle \hat{H}\rho_T \rangle_j = \sum_{k=0}^{N-1} \frac{e^{-E_k/T}E_k}{Z} \sum_{j=0}^{N-1} \vert \langle j \vert e_k\rangle \vert^2 = \langle \hat{H} \rangle \,,
\end{gather}
from which the FI for a position measurement \eqref{eq:fisherinfoposition} follows.

\section{QFI and FI for the Cartesian product of two graphs}
\label{app:qfifi_cartprod}
\subsection{Cartesian product of two graphs}
The Cartesian product $G_1 \square G_2$ of two graphs $G_1$ and $G_2$ is a graph with vertex set $V(G_1) \times V(G_2)$. Therefore, a generic vertex of $G_1 \square G_2$ is denoted by $(j,k)\in V(G_1) \times V(G_2)$ and the adjacency of vertices is determined as follows: two vertices $(j,k)$ and $(j',k')$ are adjacent if either ($j=j'$ and $k \sim k'$) or ($j \sim j'$ and $k =k'$), where the $\sim$ symbol indicates the adjacency relation between two vertices. If $G_1$ and $G_2$ are graphs on $N_1$ and $N_2$ vertices, respectively, then the Laplacian matrix of $G_1 \square G_2$ is
\begin{equation}
L(G_1 \square G_2) = L(G_1) \otimes I_{N_2}+I_{N_1} \otimes L(G_2)\,,
\label{eq:LaplCartProd}
\end{equation}
where $I_N$ denotes the $N\times N$ identity matrix. If $(E^{(1)}_1,\ldots,E^{(1)}_{N_1})$ and $(E^{(2)}_1,\ldots,E^{(2)}_{N_2})$ are the Laplacian spectra of $G_1$ and $G_2$, respectively, then the eigenvalues of $L(G_1 \square G_2)$ are 
\begin{equation}
E^{(1)}_m+E^{(2)}_n\,,
\label{eq:eigval_cartprod}
\end{equation}
with $1 \leq m \leq N_1$ and $1 \leq n \leq N_2$. Moreover, if $\vert e_m^{(1)} \rangle$ is the eigenstate of $L(G_1)$ corresponding to $E_m^{(1)}$, and $\vert e_n^{(2)} \rangle$ the eigenstate of $L(G_2)$ corresponding to $E_n^{(2)}$, then 
\begin{equation}
\vert e_m^{(1)} \rangle \otimes  \vert e_n^{(2)} \rangle
\label{eq:eigstate_cartprod}
\end{equation}
is the eigenstate of $L(G_1 \square G_2)$ corresponding to $E_m^{(1)}+E_n^{(2)}$ \cite{barik2015laplacian}.

\subsection{Quantum Fisher Information}
The Laplacian matrix $L(G)$ is the Hamiltonian of a CTQW on the graph $G_1 \square G_2$. According to the energy eigenvalues \eqref{eq:eigval_cartprod}, the partition function is
\begin{equation}
Z(G_1 \square G_2) = Z(G_1) Z(G_2)\,,
\label{eq:partfn}
\end{equation}
where $Z(G_1)$ is the partition function for a CTQW on the graph $G_1$, and $Z(G_2)$ is the partition function for a CTQW on the graph $G_2$. It follows that the expectation value of the energy is
\begin{equation}
\langle \hat{H}(G_1 \square G_2) \rangle= \langle \hat{H}(G_1) \rangle +\langle \hat{H}(G_2) \rangle\,.
\end{equation}
Moreover
\begin{align}
\langle \hat{H}^2(G_1 \square G_2) \rangle = &\langle \hat{H}^2(G_1) \rangle +\langle \hat{H}^2(G_2) \rangle\nonumber\\
&+2\langle \hat{H}(G_1) \rangle \langle \hat{H}(G_2) \rangle \,,
\end{align}
and so the QFI \eqref{eq:QFIcartprod} follows by definition \eqref{eq:qfi_def}.

\subsection{Fisher Information for position measurement}
A generic vertex of $G_1 \square G_2$ is $(j,k) \in V(G_1) \times V(G_2)$, meaning that $j \in V(G_1)$ and $k \in V(G_2)$. Accordingly, a position eigenstate in $G_1 \square G_2$ is $\vert j \rangle\otimes \vert k \rangle$. According to Eqs. \eqref{eq:eigval_cartprod}--\eqref{eq:partfn}, the Gibbs state is
\begin{align}
\rho_T(G_1 \square G_2) =\rho_T(G_1) \otimes \rho_T(G_2)\,.
\end{align} 
The probability of finding the walker in $(j,k)$ at a given temperature $T$ is
\begin{align}
p(j,k \vert T) &= \Tr \left[ \rho_T(G_1 \square G_2) \dyad{j}\otimes\dyad{k}\right]\nonumber\\
&=\sum_{m} \frac{e^{-E_m^{(1)}/T}}{Z(G_1)} \vert \langle j \vert e_m^{(1)}\rangle \vert^2 \sum_{n} \frac{e^{-E_n^{(2)}/T}}{Z(G_2)}\vert \langle k \vert e_n^{(2)}\rangle \vert^2 \nonumber\\
&=p_1(j \vert T)p_2(k \vert T)\,,
\end{align}
where $p_1(j\vert T)$ is the probability of finding the walker in the vertex $j$ of $G_1$, and, analogously, $p_2(k \vert T)$ is the probability of finding the walker in the vertex $k$ of $G_2$. Notice that $\sum_j p_1(j \vert T) = \sum_k p_2(k \vert T) = 1$. Since
\begin{align}
\partial_T p(j,k \vert T) = &\left[ \partial_T p_{1}(j\vert T)\right] p_{2}(k \vert T)\nonumber\\
&+ p_{1}(j\vert T) \partial_T p_{2}(k \vert T)\,,
\end{align}
we find that the FI \eqref{eq:fisherinfoposition} is
\begin{align}
\mathcal{F}_c(G_1 \square G_2 \vert T) &= \sum_{j} \frac{\left( \partial_T p_{1}(j\vert T)\right)^2}{p_{1}(j\vert T)} \sum_k p_{2}(k \vert T)\nonumber\\
&\quad+ \sum_ k \frac{\left( \partial_T p_{2}(k \vert T)\right )^2}{p_{2}(k \vert T)} \sum_j p_{1}(j \vert T) \nonumber\\
&\quad+ 2 \sum_{j}  \partial_T p_{1}(j \vert T) \sum_{k} \partial_T p_{2}(k\vert T)\,,
\end{align}
from which Eq. \eqref{eq:FIcartprod} follows, since $\sum_j \partial_T p_{1}(j \vert T) = \partial_T \sum_j p_{1}(j,T)=0$ and analogously $\sum_k \partial_T p_{2}(k \vert T)=0$.

\subsection{Grid graph and torus grid graph}
In this section we offer some details to assess the QFI and the FI for the grid graph and torus grid graph in Table \ref{tab:Q_FI_comparison}. In particular, we report the number of edges $M$ and the sum of the degrees squared $\sum_k d_k^2$ required to compute the QFI \eqref{eq:QFI_highT} and the FI \eqref{eq:FI_highT} in the high-temperature regime, as well as the energy level $E_1$ and its degeneracy $g_1$ required to compute the QFI \eqref{eq:lowTqfi_approx} in the low-temperature regime.

The grid graph $G_{N,N}=P_N\square P_N$ is the Cartesian product of two path graphs $P_N$, and represents a $N\times N$ square lattice with OBCs. The total number of vertices is $N^2$, while the number of edges is $M =2 N (N-1)$. There are four vertices with degree $2$ (the corners), $(N-2)$ vertices with degree $3$ on each side of square lattice, and the remaining $N^2-4-4(N-2)=(N-2)^2$ vertices have degree $4$. Hence $\sum_k d_k^2 = 4 (4N^2-7N+2)$. The path graph $P_N$ has nondegenerate energies $E_0=0$ and $E_1 = 2[1-\cos(\pi/N)]$. The grid graph has exactly the same $E_1$ but with degeneracy $g_1=2$, since, according to Eq. \eqref{eq:eigval_cartprod}, it results from the two possible combinations of $E_0$ and $E_1$ of the two $P_N$.

The torus grid graph $T_{N,N}=C_N\square C_N$ is the Cartesian product of two cycle graphs $C_N$, and represents a $N\times N$ square lattice with PBC. The total number of vertices is $N^2$, while the number of edges is $M =2 N^2$. It is $4$-regular, hence $\sum_k d_k^2 = 16 N^2$. The cycle graph $C_N$ has nondegenerate energy $E_0=0$ and $2$-degenerate energy $E_1 = 2[1-\cos(2\pi/N)]$. The torus grid graph has exactly the same $E_1$ but with degeneracy $g_1=4$, since, according to Eq. \eqref{eq:eigval_cartprod}, it results from the four possible combinations of $E_0$ and $E_1$ of the two $C_N$.

\section{Exact QFI for the complete bipartite graph}
\label{app:QFIcbg}
The energy spectrum of the complete bipartite graph $K_{N_1,N_2}$ consists of only four energy levels (see Sec. \ref{sec:CBG}). This allows us to obtain the QFI at all the temperatures from Eq. \eqref{eq:qfi_def}
\begin{align}
    \mathcal{F}_q(T) = &\frac{e^{-2 (N_1 + N_2)/T}}{Z^2T^4}
     \left\lbrace N_1^2 e^{N_1/T} \left[ (N_1-1)+e^{2N_2/T} (N_2-1)\right]\right.\nonumber\\
     &+N_2^2 e^{N_2/T} \left[ e^{2N_1/T} (N_1-1)  + (N_2-1)\right]\nonumber\\
   &+ e^{(N_1 + N_2)/
    T} \left[N_1^3 (N_2-1 ) - N_2^2(N_2-2 )\right.\nonumber\\
    &\left.\left. + N_1 N_2^2 (N_2+1)- 
      N_1^2 (2 N_2^2-N_2-2 )\right]\right\rbrace\,,
\end{align}
where $Z=1 + (N_2-1)e^{-N_1/T}+ (N_1-1)e^{-N_2/T} + e^{-(N_1 +N_2)/T}$. For the star graph $S_N$, which is the complete bipartite graph $K_{1,N-1}$, the spectrum reduces to three energy levels, and the resulting QFI is
\begin{align}
    \mathcal{F}_q&(T)\nonumber\\
    &= \frac{e^{-(N+1)/
  T} \left[e^{N/T} (N-2) + (N-2) (N-1)^2 + e^{1/T} N^2\right]}{T^4 \left[(1 + (N - 2) e^{-1/T} + e^{-N/T}\right]^2}\,.
\end{align}

\bibliography{BibTopologyThermometry}
\end{document}